# Key parameters affecting the performance of fractured geothermal reservoirs: a sensitivity analysis by thermo-hydraulic-mechanical simulation

Saeed Mahmoodpour[1], Mrityunjay Singh[1], Aysegul Turan[1], Kristian Bär[1], Ingo Sass[1,2]

[1] Technische Universität Darmstadt, Institute of Applied Geosciences, Group of Geothermal Science and Technology, Schnittspahnstrasse 9, 64287 Darmstadt, Germany

[2] Darmstadt Graduate School of Excellence Energy Science and Engineering, Otto-Berndt-Strasse 3, 64287 Darmstadt, Germany

*Correspondence to*: Saeed Mahmoodpour (saeed.mahmoodpour@tu-darmstadt.de)

**Abstract**

A discretely fractured reservoir has the capability to support and sustain long term geothermal energy extraction operations. The process of injecting cold water to extract hot water from a fracture reservoir results in thermal and poroelastic stresses in the rock matrix. Therefore, these thermo-hydro-mechanical (THM) mechanisms govern the efficiency of an enhanced geothermal system (EGS) operation. However, THM mechanisms are governed by various rock and fluid parameters as well as on the initial and boundary conditions of the model set-up. In this paper, we have identified and analyzed 22 such parameters. Due to this large number of involved parameters, it is difficult to accurately estimate the relative importance of individual parameters. To address this issue, we have performed an extensive sensitivity analysis for the purpose of the Horizon 2020 project, Multidisciplinary and multi-contact demonstration of EGS exploration and Exploitation Techniques and potentials (H2020 MEET) using the design of the experiment method and identified key parameters influencing three key outcomes of the simulation: thermal breakthrough time, mass flux, and overall energy recovery. We found that for a given discrete fracture network, the fracture aperture, the rock matrix permeability, and the wellbore radius are the most influential parameters controlling the thermal breakthrough time, mass flux, and overall energy recovery.

## 1. Introduction

The world is facing a carbon negative energy crisis. Hydrocarbon based energy sources has projected a detrimental impact on the climate change due to anthropogenic emission of carbon dioxide into the atmosphere. Carbon dioxide geosequestration in deep saline reservoirs is the most effective method to mitigate global warming (Mahmoodpour et al. 2019, 2020; Singh et al. 2020, 2021). However, alternative green energy resources, such as geothermal energy, are globally becoming lucrative (Lu 2018; Jolie et al. 2021). Geothermal energy is efficient, environment friendly, sustainable, all-weather available, and a renewable resource of energy (Tester et al. 2007; Chamorro et al. 2012; Li et al. 2015; Pandey et al. 2017; Will et al. 2017; Singh et al. 2020). Furthermore, it is a low carbon foot-print resource and has the potential to provide energy for 217 million years as per year 2012 global energy consumption rate (Lu 2018). The earth has $1.3 \times 10^{27}$ J of energy stored within 10 km of from its surface (Lund 2007). Jain et al. (2015) estimated that available land surface for EGS in Germany could provide 4155 TWh of electrical energy in one year, equivalent to seven times the total energy produced in the year 2011. This energy is harnessed using water either through naturally available hydrothermal resources (Parisio et al. 2019) or by engineering fractured reservoirs (Harlow and Pracht 1972). Fractured reservoirs features low porosity and permeability with a high geothermal gradient. Therefore, an enhanced geothermal system (EGS) is established by creating artificial fractures in fractured reservoirs where cold heat carrier fluid (most commonly water) flows through the fractures and hot fluid is extracted. Fracture networks in fractured reservoirs show intricate geometrical characteristics, thereby making it difficult to accurately estimate the overall geothermal energy extraction. At the same time injection of cold water and production of hot water causes thermo-poroelastic stress in fractured reservoirs . Therefore, the understanding of coupled thermo-hydro-mechanical (THM) processes in a discretely fractured rock is necessary for prediction of energy extraction capability and unwanted microseismic activities.

**THM process**

Fractured reservoirs are complex geological settings and, when developed for enhanced geothermal energy extraction, even become complicated due to the inclusion of discrete heterogeneity, introduced by including fractures and faults (Cacace and Jacquey et al. 2017). To investigate the EGS operation in a fractured reservoir, it is necessary to perform numerical modeling of heat transfer, mass transfer, chemical species transport, and geomechanical deformation in a fully coupled manner. The scope of this work is limited to THM coupled processes involved in EGS operation.

Coupled THM models for geothermal energy extraction from fractured reservoirs have been rigorously investigated in last five decades. Harlow and Pracht (1972), McFarland (1975), Abe et al. (1976a, b), and Kohl et al. (1995) modeled THM processes to investigate hydraulic fracturing on energy recovery using a penny shaped fracture. Bažant and Ohtsubo (1978) worked on numerical modeling for temperature change impact on crack thickness associated with THM process, whereas DeTeaux et al. (1996) investigated the change in rock joint properties. Further, THM mechanism occurring during geothermal energy extraction was sophisticated by Hicks et al. (1996), Ghassemi et al. (2003), Wanatabe et al. (2010), Ghassemi and Zhou (2011), Guo et al. (2016), Cao et al. (2016), and Pandey et al. (2017) for single fracture reservoir.

The aperture variation negatively impacts the lifetime operation of an EGS project. Variations in normal and shear stresses acting on the fracture surfaces, due to THM processes, causes fracture aperture variations which has a significant impact on fluid transport and the overall energy recovery (Rutqvist et al. 2005). Cacace and Jacquey et al. (2017) performed THM numerical simulations in a geothermal doublet system with fractures present at the injection-production wells, and found that heat extraction as well as pore pressure may induce thermal stress and effective stress of the rock, respectively. They also proposed that effective stress can change the fracture permeability and ultimately control the overall heat recovery. Assuming the inherent stress benefits in creating horizontal wells while hydraulic fracturing, Sun et al. (2018) proposed a dual horizontal well EGS model to enhance the contact area. However, their model could not demonstrate a commercial scale energy efficiency. Considering the fracture system as a high permeability region in a fractured reservoir system, Yin et al. (2018) found that excessive pressure depletion may damage the formation during hot water extraction and overpressure may uplift the fractured reservoirs in the early period, whereas thermal contraction subside it in the later period. They also observed that overpressure propagation is much faster than cold front propagation along the injection-production axis. At the same time, Vik et al. (2018) suggested that matrix contraction has a favorable impact on heat extraction as it increases the fracture aperture and supports flow pathway evolution. However, they also reported that an increase in fracture aperture creates channel formation between the wells. These channels cause short circuiting which further reduces the time for cold water to exchange heat from the rock, and therefore less heat production is observed. Salimzadeh et al. (2018) modeled a THM processes by sequentially coupling the TH (Thermo-Hydro) process with thermoporoelastic compression for a single fracture in a fractured reservoir. They noted that fractured reservoirs show a smaller thermal expansion coefficient than water and, due to a very small magnitude of thermal diffusivity of the rock, pore pressure dissipates even in very low-permeability rocks. Therefore, the thermal expansion coefficient measured at undrained condition scan overestimate the volumetric strain, whereas it underestimates at drained conditions. They suggested to use an effective thermal expansion coefficient from drained and undrained conditions for EGS operations. Recently, Aliyu and Archer (2020a, b) performed THM numerical simulations to optimize energy recovery from fractured reservoir. They optimized the energy extraction based on the number of fractures and fracture spacing, as well as for wellbore position, respectively. In another study, Aliyu and Archer (2020c) found that for a single fracture system, THM coupled processes indicate greater variation in fracture aperture compared to TH coupled processes. For a set of fractures at specified constant spacing, Aliyu and Archer (2021) observed that the geomechanical aspect of fracture aperture, permeability, and reservoir stiffness governs the EGS performance.

**DFN-THM**

THM modeling of discretely fractured reservoir for EGS operation is a relatively new modeling approach. For a three-dimensional fracture network undergoing THM process, Safari and Ghassemi (2015) observed that fluid injection and extraction develops gradual shear in fractures. On a continuous injection of fluid, they found that stress intensity favors the fracture propagation in shear and tensile modes that enhances the reservoir surface area to increase seismicity. Sun et al. (2017) performed numerical simulations for THM processes involved in EGS, based on a two-dimensional discrete fracture network model. Coupling THM processes, Ghassemi and Zhou (2011) and Salimzadeh et al. (2017c) found that fluid leak-off in the rock matrix enhances the heat extraction and delays the cold water breakthrough at the production well. Concurrently, Salimzadeh et al. (2017a,b) observed that fluid pressure increases due to leak-off causing rock expansion and evolution of back-stress. Stress relaxation caused by interacting two fractures leads to an increase in the fracture aperture. The increase in fracture aperture therefore increases the fluid velocity through the fractures and reduces the thermal drawdown temperature. Later, Salimzadeh et al. (2018) estimated that fluid leak-off is possible for a permeability greater than $10^{-18}$ m$^2$. Zhang et al. (2019) modeled THM processes for a discrete fracture network (DFN) in a reservoir and found that with a continuous extraction of heat, fluid viscosity increases in the vicinity of the production well which further decreases the fluid velocity. They further found that the cold front changes the effective stress field. For a 20-year duration of fluid extraction, they observed a variation in fracture permeability, resulting in an increased outflow and faster temperature depletion at the production well. However, characteristic geometry of the fracture network enhances the flow around the tortuous fractures enhancing the heat transfer which ultimately slows down the thermal breakthrough. Thereafter, Yuan et al. (2020) found that the increase in permeability is due to the impact of coupled injection-driven cooling and fluid pressure. They proposed that the temperature drop reactivates natural fractures under hydro-shearing mechanism, whereas fluid pressure supports shear failure and assists in the stimulation zone expansion. For a three-dimensional fracture network undergoing hydro-mechanical interaction, Paluszny et al. (2020) observed that with the growth in the fracture network, small variation in the fracture geometry causes larger alteration in the position and dimension of the primary channels.

**Sensitivity study**

Fracture complexity, mechanical conditions, and fluid properties govern the overall THM process in a fractured reservoir. Hofmann et al. (2016) reviewed the controlling parameters of fracture complexity and summarized that natural discontinuity, in-situ stresses, rock mechanical properties, and hydraulic reservoir properties are the principal natural factors. Among the engineered factors, fluid volume and treatment durations, flow rates, various injection schemes, stimulation mechanisms and completion design have considerable significance. Aliyu and Chen (2017) performed sensitivity analysis for thermo-hydro processes in a geothermal reservoir to estimate the production temperature. They found that fluid injection temperature is the most important parameter influencing the production temperature, whereas reservoir temperature drop was primarily governed by injection pressure. Furthermore, they observed that naturally occurring parameters indicate stable fluid production temperature, unlike human-controlled parameters demonstrating an unstable temperature distribution. Later, Asai et al. (2019) optimized the flow rate to maximize the heat extraction efficiency from EGS. For a 30 year operational period, they proposed that an exponential fluid flow scheme is the most optimized strategy and the heat extraction per unit of water is inversely proportional to the total water circulated. From literature we have found that at least 22 parameters govern these coupled physical processes during an EGS operation. However, it is a tedious task to find which of these parameter has the most critical role in controlling THM processes. Therefore, in this paper we have performed a sensitivity analysis to identify and categorize these parameters. Sensitivity analysis for the parameters involved in THM process during EGS operation is scattered and unorganized in the literature. This study fills the gap in the literature by presenting the most important findings of these sensitivity and optimization studies. Alternatively, machine learning approaches can be adopted to predict the critical parameters with similar accuracy (Pandey and Singh 2021) but at the expense of higher computational cost.

The initial reservoir temperature is the primary factor in determining a rough estimation of the total energy extraction capacity based on the geothermal gradient. Interestingly, Wang et al. (2019) observed that water-rock heat convection in a fracture is highly intensive, compared to heat conduction in rocks at different temperatures. Zhang et al. (2019) observed that the variation in the temperature distribution leads to thermal conductivity change within the reservoir which ultimately changes the fluid flow and stress distribution. They further found that the reservoir heat capacity evolves with thermal conductivity.

Fluid transmission is controlled by the thermophysical properties of fluid, wellbore radius, and the injection rate. Li et al. (2015) found that the wellbore radius is the most important factor during fluid flow in an EGS. Cold fluid propagation from the injection to production well increases the fluid viscosity which subsequently decreases the fluid flow rate (Guo et al. 2019). Cao et al. (2016) observed that an injection temperature drop and an injection pressure increase develops negative effective stress, leading to enhanced porosity and permeability. Comparing a hydro-mechanical coupled experiment with a THM numerical model, Yuan et al. (2020) found that an experimentally measured injection rate increases drastically, unlike to numerical model where it increases gradually. They proposed an explanation linked to an early failure of fractures near the injection well. Moreover, there was a drop in the early period injection rate in the measured injection rate due to instability in the formation.

The mean fracture aperture is a function of normal contact stress for the fracture surfaces present in the partial contact. The classical Barton-Bandis model (Bandis et al. 1983; Barton et al. 1985) is popular in modeling the fracture aperture for this condition. Izadi and Elsworth (2014) found that during an EGS operation involving coupled THMC mechanisms, resulting in permeability evolution, is slower with time for a low density fracture network. Fracture connectivity plays a crucial role within the lifetime of the project and the overall energy recovery. Han et al. (2019) observed that undeveloped fracture network allow higher injection pressure, thereby a higher heat extraction ratio is possible at the expense of a reduced total project lifespan and thermal breakthrough. However, they found that a fully connected fracture network causes thermal short-circuiting and faster thermal breakthrough. They suggested that avoiding a preferential channel whilst creating a fracture network can support higher heat extraction. Fracture connectivity is more significant than fracture spacing for energy extraction (Xin et al. 2020).

Processes controlled by fracture aperture varies during the EGS operation for single fracture models (Wang et al. 2016; Vik et al. 2018) and DFN models (Guo et al. 2016). Wang et al. developed a correlation for thermally induced fracture aperture variation. Guo et al. (2016) observed that fracture aperture heterogeneity non-uniformly decrease the rock matrix temperature. Due to this, the thermal stresses causes preferential flow and therefore, faster depletion in heat production is possible in fractured reservoirs. They proposed that tortuous preferential flow paths may increase the longevity of the heat extraction. Salimzadeh et al. (2018) proposed that a maximum fracture aperture occurs behind the injection well, instead of at the injection well, due to stress redistribution. Similarly, Qarinur et al. (2020) found that heat contraction and increase in fluid pressure reduces the fracture contact surface stress which further increases the fracture aperture.

The fractured reservoir equivalent permeability depends on the aperture distribution model (Bisdom et al. 2016), where the equivalent permeability is necessary for Darcy flow model simulations. Salimzadeh et al. (2018) found that rock permeability has a minor effect on temperature distribution due to the dominant conductive heat transfer mechanism in the system. Smaller magnitude of Young's modulus decreases the thermally induced stress in the matrix, resulting in a smaller fracture aperture. Therefore, the EGS operation longevity increases with a lower Young's modulus (Vik et al. 2018).

Zimmerman (2000) found that the minimum value of Biot coefficient is $3\varphi/(2+\varphi)$, and is a decreasing function of porosity (Tan and Konietzky 2017). The volumetric contraction is small for smaller Biot coefficient due to the variation in rock matrix pressure and decrease in fracture aperture. For granites a realistic value will be $\alpha=0.2$ due to its extremely low permeability (Salimzadeh et al. 2018). In this text, we have used a higher value of Biot's coefficient due to higher reservoir permeability.

Improper or unsustainable generation of fractured reservoirs may damage the reservoir and impact the energy recovery process. The difficulties may arise including reservoir dryness, pressure drawdown, early thermal drawdown, and silica scaling (Kaya et al. 2011; Rivera Diaz et al. 2016). From field experiences such as Soultz-sous-Forêts, Fenton Hill, Rosemanowes, and Paralana, achieving optimal flow rate and short-circuiting were the primary challenges (Parker 1999; Brugger et al. 2005; Evans 2005; Genter et al. 2010; Haris et al. 2020). To avoid the cost and time consumption involved, numerical simulations are the most reliable and effective method to accurately predict the energy recovery from a fractured reservoir (Hayashi et al. 1999; O'Sullivan et al 2001). However, a large number of parameters affect the THM simulation of the fractured geothermal reservoirs. Due to the high costs or measurement limitations, it is important to know the relative importance of the key parameters.

Here we aim to use the well-known design of a experiment-based statistical approach to examine the sensitivity of the involved parameters during heat extraction from randomly fractured geothermal reservoirs through a fully coupled THM simulation. The fractured reservoir for this study is characterized as a linear elastic, homogeneous, and isotropic material. The present study will establish a quick guideline tool in producing the optimal designs for fractured reservoir-EGS systems under the Horizon 2020 project, Multidisciplinary and multi-contact demonstration of EGS exploration and Exploitation Techniques and potentials (H2020 MEET). We numerically modeled geothermal energy recovery from fractured reservoir and associated fluid flow, thermo-poroelastic stress evolution, and heat transport mechanisms. Further, we present a sensitivity study based on these numerical simulations and identified the comparatively important parameters.

## 2. Methods

### 2.1 Model schematic, initial and boundary conditions

A Discrete fracture network (DFN) map from Otsego County in New York, USA is taken as a basis during simulation (Mahmoodpour and Masihi 2016). For the sake of the computational efficiency, a two-dimensional planar model (1000 m × 600 m) with embedded random fractures is adopted to investigate the thermo-poroelastic responses during heat extraction (as shown in Figure 1), as it is unfeasible to reproduce the actual three dimensional fracture system of the buried reservoir. It is also noticeably profitable to model all fractures as interior boundaries, because it eliminates the need to create a geometry with a high aspect ratio. All fractures within the domain are regarded as internal boundaries, implicitly considering the mass and energy exchange between porous media and fractures. We have constrained the displacement in all normal directions. All boundaries of the modeled domain are defined as no flow for both fluid and heat transmission.

### 2.2 Governing equations

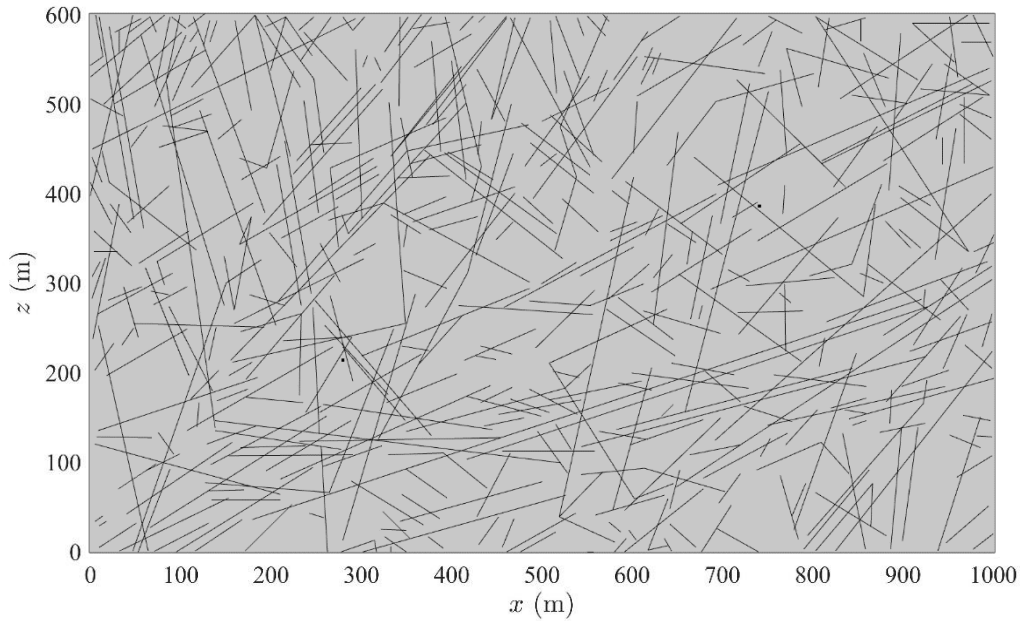

*Figure 1: Geometry of the model*

#### 2.2.1 Mass conservation

The effect of changes in pore volume and fluid temperature on mass transport is incorporated into the mass balance equation. The final governing equation describing fluid flow in porous media can be written as follows:

$$\rho_1 C_{1m} \frac{\partial p}{\partial t} - \rho_1 C_{2m} \frac{\partial T}{\partial t} + \rho_1 C_{3m} \frac{\partial \varepsilon_V}{\partial t} = \nabla \cdot \left(\frac{\rho_1 k_m}{\mu} \nabla p\right) \qquad [1]$$

where p, T, $\varepsilon_V$ are the fluid pressure, fluid temperature, and pore volumetric strain in porous media respectively; $C_{1m} = \phi_m S_1 + (1 - \phi_m) S_m$ represents the storage coefficient of porous media; $C_{2m} = \alpha_m(\phi_m \beta_1 + (1 - \phi_m)\beta_m)$ represents the thermal expansion coefficient of porous media; $C_{3m} = \alpha_m$ represents Biot's coefficient of porous media; $\phi_m$ is the reservoir porosity; $S_m$ and $S_1$ represent the storage coefficients of the fluid and rock matrix; $\beta_1$ and $\beta_m$ represent the thermal expansion coefficients of the fluid and rock matrix; $\rho_1$ and $\mu$ are the density and dynamic viscosity of the fluid; and $k_m$ is the pressure-dependent pore permeability, respectively.

Preexisting fractures in porous media can be regarded as internal boundaries. Comparing the fracture width (in millimeter) with the fracture length (in meter), the flow along the fracture width direction can be ignored. The tangential derivatives are used to define the flow along the internal fracture by

$$\rho_1 C_{1f} e_h \frac{\partial p}{\partial t} - \rho_1 C_{2f} e_h \frac{\partial T}{\partial t} + \rho_1 C_{3f} e_h \frac{\partial \varepsilon_V}{\partial t} = \nabla_T \cdot \left( \frac{e_h \rho_1 k_f}{\mu} \nabla_T p \right) + n. Q_m \qquad [2]$$

where p, T, $\varepsilon_V$ are the fluid pressure, fluid temperature, and pore volumetric strain in the fracture respectively; $C_{1f} = \phi_f S_1 + (1 - \phi_f) S_{mf}$ represents the storage coefficient of the fracture; $C_{2f} = \alpha_f(\phi_f \beta_1 + (1 - \phi_f)\beta_f)$ represents the thermal expansion coefficient of the fracture; $C_{3f} = \alpha_f$ represents Biot's coefficient of the fracture; $\phi_f$ is the fracture porosity; $S_f$ represents the storage coefficients of the fracture; $\beta_f$ represents the thermal expansion coefficient of the fracture; $e_h$ is the hydraulic aperture between the two fracture surfaces; $k_f$ is the stress-dependent fracture permeability; $nQ_m = n.(-\frac{\rho k_m}{\mu \nabla p})$ is the mass flux exchange between porous media and the fracture; and $\nabla_T$ denotes the gradient operator restricted to the fracture's tangential plane.

### 2.2.2 Energy balance

In the injection well, the diameter of the well is small and can be represented by a line. Therefore, the linear heat source can be calculated by the product of injected water mass times specific heat capacity of water times the temperature difference between local rock temperature and injection temperature.

The local thermal non equilibrium (LTNE) theory is adopted to simulate the heat exchange between the rock matrix and the flowing fluid. For rock matrix, the energy transfer process is mainly dominated by the heat conduction and the heat exchange between pore fluid. The governing equation for rock matrix is satisfied by

$$(1 - \phi_m)\rho_m C_{p,m} \frac{\partial T_m}{\partial t} = \nabla \cdot \left((1 - \phi_m)\lambda_m \nabla T_m\right) + q_{ml}(T_l - T_m) \qquad [3]$$

where $T_m$ and $T_l$ are the matrix and fluid temperatures in porous media, respectively; $\rho_m$ is the density of the rock matrix; $C_{p,m}$ is the specific heat capacity of the rock matrix; $\lambda_m$ is the heat conductivity of the rock matrix; and $q_{ml}$ represents the rock matrix-pore fluid interface heat transfer coefficient. From the domain's point of view, the ingoing heat flux is received from the thin fracture. Conversely, the outgoing heat flux $n. q_m$ leaves the domain and is received by the adjacent fracture in the source term

$$(1 - \phi_f) e_h \rho_f C_{p,f} \frac{\partial T_m}{\partial t} = \nabla_T \cdot \left((1 - \phi_f) e_h \lambda_f \nabla_T T_m\right) + e_h q_{fl}(T_l - T_m) + n. q_m \qquad [4]$$

where $T_m$ and $T_l$ are the matrix and fluid temperatures in the fracture, respectively; $\rho_f$ is the density of the fracture; $C_{p,f}$ is the specific heat capacity of the fracture; $\lambda_f$ is the heat conductivity of the fracture; and $q_{fl}$ represents the rock fracture-fluid interface heat transfer coefficient, related to the fracture aperture; and $n. q_m = n.(-(1 - \phi_m)\lambda_m \nabla T_m)$ represents the heat flux exchange of the solid between rock matrix and the fracture. For pore fluid, the contribution of heat convection should be incorporated into the aforementioned energy balance equation

$$\phi_m \rho_l C_{p,l} \frac{\partial T_l}{\partial t} + \phi_m \rho_l C_{p,l} u_m. \nabla T_l = \nabla \cdot (\phi_m \lambda_l \nabla T_l) + q_{ml}(T_m - T_l) \qquad [5]$$

where $C_{p,l}$ is the heat capacity of the fluid at a constant pressure; $\lambda_l$ is the heat conductivity of the fluid; and $u_m = -\frac{k_m \nabla p}{\mu}$ is the Darcy's rate in theporous media.

The heat flux coupling relationship of the fluid between the domain and the fracture is satisfied by

$$\phi_f e_h \rho_l C_{p,l} \frac{\partial T_l}{\partial t} + \phi_f e_h \rho_l C_{p,l} u_f . \nabla_T T_l = \nabla_T . (\phi_f e_h \lambda_l \nabla_T T_l) + e_h q_{fl}(T_m - T_l) + n.q_l \qquad [6]$$

where $u_f = -\frac{k_f \nabla_T p}{\mu}$ is the flow rate in fractures; the heat flux $n.q_l = n.(-\phi_l \lambda_l \nabla T_l)$ denotes the heat exchange of the fluid between porous media and the fracture.

Meanwhile, the temperature-dependent fluid thermodynamic properties are incorporated into the aforementioned mass and energy conservation equations, to account for the bidirectional HT coupling. The change in fluid density, viscosity, heat conductivity, and specific heat capacity with the temperature are described by some empirical formulas (See Eq. [10-14]).

### 2.2.3 Stress

Based on the linear elasticity, effective stress, and thermal stress theory, the thermo-poroelastic equation describing the stress-strain relationship of porous media is described as follows

$$\sigma_{ij} = 2G\varepsilon_{ij} + \lambda tr\varepsilon \delta_{ij} - \alpha_p p \delta_{ij} - K'\beta_T T \delta_{ij} \qquad [7]$$

where $\sigma_{ij}$ is the total stress; G and $\lambda$ are Lame's constants; tr is the trace operator; the third and fourth terms on the right hand of Eq. [7] represent the poroelastic effect and thermo-elastic effect, respectively; $K' = \frac{2G(1+\nu)}{3(1-2\nu)}$ is the bulk modulus of the drained porous media; $\beta_T = \phi_l \beta_l + (1-\phi_m)\beta_m$ is the volumetric thermal expansion coefficient of porous media; and $\delta_{ij}$ is the Dirac dealt function; $\alpha_p$ represents the Biot's coefficient. $\sigma_{eff}^{ij} = \sigma_{ij} + \alpha_p p \delta_{ij}$ is the effective stress. In the present context, the tension is conventionally indicated by a positive notation.

Combining the equilibrium equations, the deformation equation of porous media can be derived as follows

$$Gu_{i,jj} + (G+\lambda)u_{j,ji} - \alpha_p p_{,i} - K'\beta_T T_{,i} + f_i = 0 \qquad [8]$$

where $f_i$ denotes the external body force.

The change in effective stress has a significant effect on the permeability of the fractured reservoir. During heat mining, the injection and extraction of the fluid will result in the variation of the pore pressure. In addition, the large temperature difference between the injected cold fluid and the hot rock mass can induce thermal tensile stress. The thermo-poroelastic effect will cause the fracture closure, opening, or shear slip, which can affect mass and heat transport during heat extraction.

A model describing the opening and closure processes of the stress-dependent fracture has been developed by Barton and Bandis (Bandis et al. 1983; Barton et al. 1985). The change in the initial aperture of the fracture under in-situ stresses can be expressed as follows

$$\Delta e_n = \frac{e_0}{1 + 9\frac{\sigma_{eff}^n}{\sigma_{nref}}} \qquad [9]$$

where $e_0$ is the initial aperture of the fracture; $\sigma_{eff}^n$ is the effective normal stress acting on the fracture surface; and $\sigma_{nref}$ is the effective normal stress required to cause 90% reduction in fracture aperture.

### 2.2.4 Fluid properties

The thermophysical properties of water such as dynamic viscosity, specific heat capacity, density, and thermal conductivity as a function of temperature are presented below:

Dynamic viscosity:

$$\mu = 1.38 - 2.12 \times 10^{-2} \times T^1 + 1.36 \times 10^{-4} \times T^2 - 4.65 \times 10^{-7} \times T^3 + 8.90 \times 10^{-10} \times T^4 - 9.08 \times 10^{-13} \times T^5 + 3.85 \times 10^{-16} \times T^6 \quad (273.15 - 413.15\ K) \quad [10]$$

$$\mu = 4.01 \times 10^{-3} - 2.11 \times 10^{-5} \times T^1 + 3.86 \times 10^{-8} \times T^2 - 2.40 \times 10^{-11} \times T^3 \quad (413.15 - 553.15\ K) \quad [11]$$

Specific heat capacity:

$$C_p = 1.20 \times 10^4 - 8.04 \times 10^1 \times T^1 + 3.10 \times 10^{-1} \times T^2 - 5.38 \times 10^{-4} \times T^3 + 3.63 \times 10^{-7} \times T^4 \quad [12]$$

Density:

$$\rho = 1.03 \times 10^{-5} \times T^3 - 1.34 \times 10^{-2} \times T^2 + 4.97 \times T + 4.32 \times 10^2 \quad [13]$$

Thermal conductivity:

$$\kappa = -8.69 \times 10^{-1} + 8.95 \times 10^{-3} \times T^1 - 1.58 \times 10^{-5} \times T^2 + 7.98 \times 10^{-9} \times T^3 \quad [14]$$

We modeled THM process in the commercial software COMSOL Multiphysics, version 5.5. Using finite element methods, COMSOL Multiphysics solves general-purpose partial differential equations. Even though, COMSOL Multiphysics has shown excellent consistency in modeling THM process during EGS operations (Aliyu and Archer 2020c, 2021), we will validate our THM model against existing analytical solutions.

## 3. Results and Discussions

In this section we present numerical simulation results for the fully coupled THM processes involved in geothermal energy extraction from a fractured reservoirFirst, we will validate our numerical model with a soil thermal consolidation model, existing in literature (Bai, 2005). Further, we will demonstrate the coupled impact of thermoelasticity and poroelasticity on fluid flow and heat transfer between a doublet for discretely fractured networks in the fractured reservoir, and assess the long term energy extraction performance. . Further, we performed sensitivity analysis employing the Plackett-Burman method and found that 48 compositional reservoir simulations are required to evaluate the crucial parameters having the greatest impact on output. Lastly, we will use these numerical simulations into a regression analyses to find the most sensitive parameters impacting the results obtained from the simulations.

### 3.1. Validation of the model

*Table 1: Magnitudes of the material parameters used for a validation model.*

| Parameter | Value | Unit |
|---|---|---|
| Fluid density | 1000 | $kg/m^3$ |
| Solid density | 2600 | $kg/m^3$ |
| Porosity | 0.4 | - |
| Hydraulic conductivity | $1 \times 10^{-9}$ | m/s |
| Thermal conductivity | 0.5 | $W/m \times K$ |
| Specific heat capacity of the fluid | 4200 | $J/kg \times K$ |
| Specific heat capacity of the soil | 800 | $J/kg \times K$ |
| Elastic modulus of the soil | 60 | MPa |
| Poisson's ratio | 0.4 | - |
| Thermal expansion coefficient | $3 \times 10^{-7}$ | - |

| Biot-Willis coefficient | 1.1 | - |
| Coefficient of compressibility | $1.1 \times 10^{-10}$ | 1/Pa |

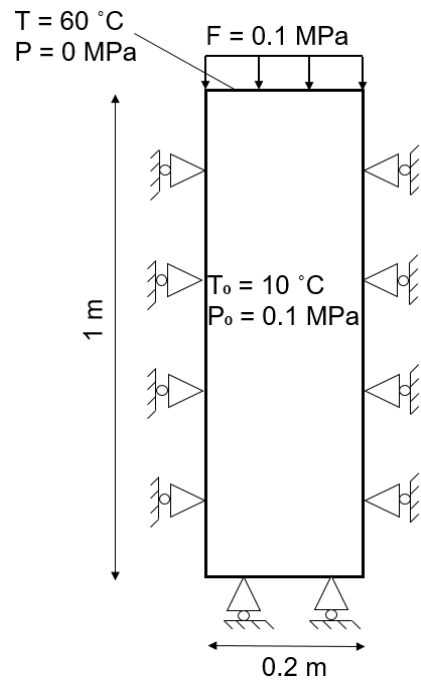

*Figure 2: Schematic geometry of a two-dimensional soil thermal consolidation model.*

We have adopted the THM coupling approach as demonstrated by Bai (2005). A schematic geometry of the soil thermal consolidation is shown in Figure 2. The dimension of the considered two-dimensional box model is 1.0 m × 0.2 m and the model is considered to be present at 10 $^O$C. Both the initial pore pressure and external pressure are 100 kPa, and 0 kPa pore pressure at 60 $^O$C is applied at the top boundary. Displacement is not allowed in the lateral and bottom boundaries, as these boundaries are assumed impermeable and insulated. Table 1 gives the material parameters used to validate the thermal consolidation problem. Figure 3 shows the comparison between analytical results obtained from Bai (2005) and numerically obtained results. It is clearly evident that displacement (see Figure 3(a)), pressure (see Figure 3(b)), and temperature (see Figure 3(c)) agrees to be in good confidence with the analytical results at different positions. Based on these results, it is obvious that our model accurately predicts the coupled THM mechanisms. This indicates the validity of our numerical approach for the case of a thermo-poroelastic coupling problem under consideration. Therefore, the model proposed in this paper can be used to reveal the thermo-poroelastic responses of geothermal reservoirs due to cold fluid injection during heat extraction.

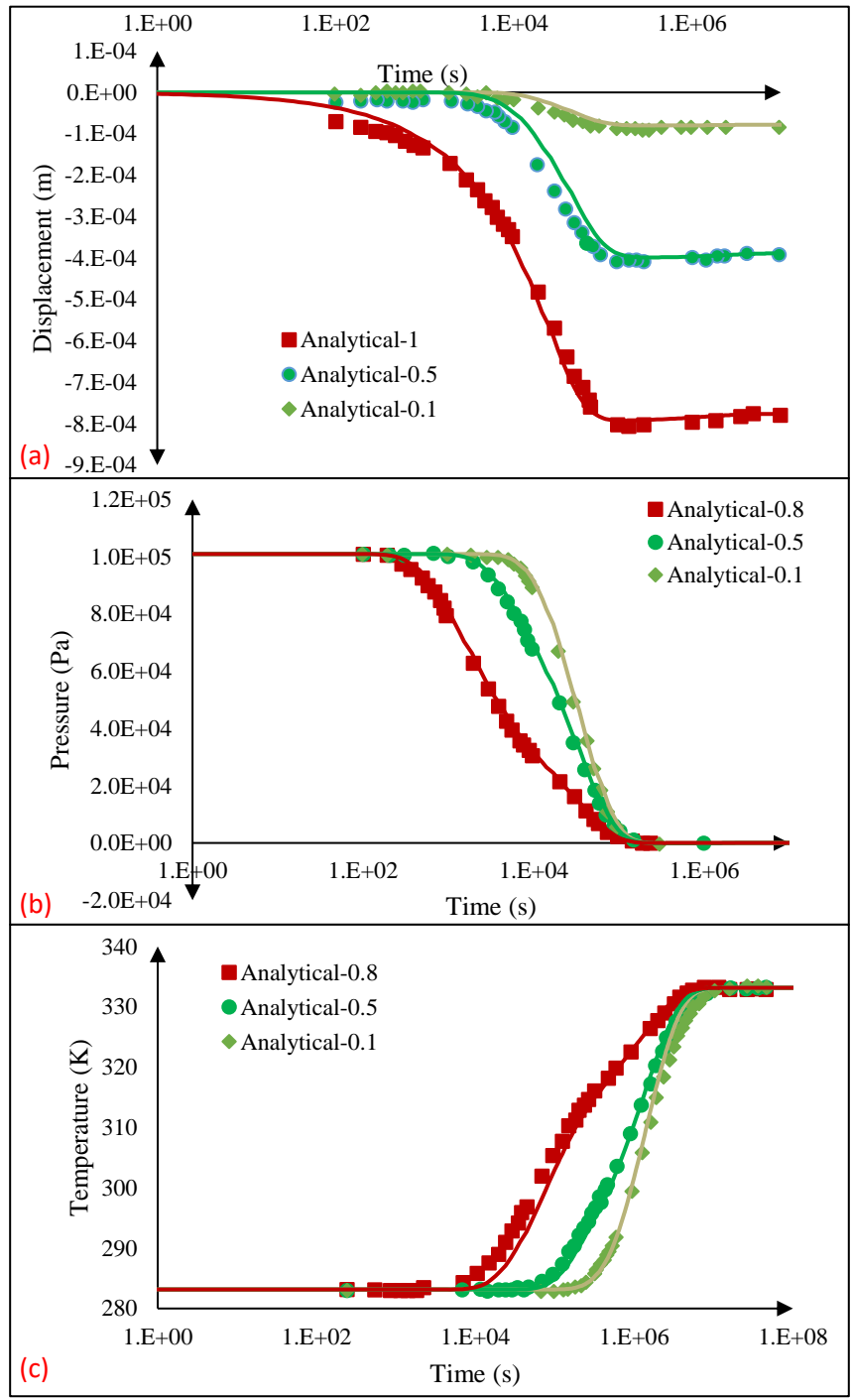

*Figure 3: (a) Displacement, (b) pressure, and (c) temperature, comparison of a thermal consolidation model from analytical solution with numerical solution at three different positions: 0.1 m, 0.5 m, and 1 m.*

### 3.2. THM modeling of randomly fractured reservoir

A discretely fractured reservoir is selected for numerically estimating the geothermal energy extraction potential by injecting cold water and producing hot water. The parameters of this representative case are listed in Table 2 where -1 indicates the lower value and 1 indicates the higher value for the corresponding parameter in Table 3.

*Table 2: Parameters used to investigate the long-term EGS potential from fractured reservoir.*

| Para-meter | E | v | ρr | S1 | S2 | Pi | Pj | φr | kr | φf | f | Ap | σs | wr | λr | λf | Cr | Cf | Ti | α | β | Tj |
|---|---|---|---|---|---|---|---|---|---|---|---|---|---|---|---|---|---|---|---|---|---|---|
| value | -1 | -1 | 1 | 1 | 1 | 1 | -1 | 1 | 1 | 1 | 1 | 1 | -1 | -1 | -1 | -1 | 1 | -1 | -1 | -1 | -1 | 1 |

*Table 3: Parameters used to investigate the sensitivity study.*

| Symbol | Parameter | Range |
|---|---|---|
| E | Young's modulus | 20 GPa − 40 GPa |
| v | Poisson's ratio | 0.2 – 0.3 |
| ρr | Rock density | $2500\,\frac{kg}{m^3}$ - $2800\,\frac{kg}{m^3}$ |
| S1 | Horizontal stress | 30 MPa – 50 MPa |
| S2 | Vertical stress | 30 MPa – 50 MPa |
| pi | Initial pressure | 20 MPa – 30 MPa |
| pj | Injection pressure | 50 MPa – 60 MPa |
| φr | Rock porosity | 0.05 – 0.2 |
| kr | Rock permeability | 10 mD – 50 mD |
| φf | Fracture porosity | 0.3 – 0.5 |
| f | Fracture roughness | 1 – 2 |
| Ap | Fracture aperture | 0.1 mm – 0.2 mm |
| σs | Closure stress | 100 MPa – 150 MPa |
| wr | Wellbore radius | 0.1 m – 0.2 m |
| λr | Rock thermal conductivity | $2\,\frac{W}{m \times K} - 3\,\frac{W}{m \times K}$ |
| λf | Fracture thermal conductivity | $1.5\,\frac{W}{m \times K} - 2\,\frac{W}{m \times K}$ |
| Cr | Rock specific heat capacity | $800\,\frac{J}{kg \times K} - 1000\,\frac{J}{kg \times K}$ |
| Cf | Fracture specific heat capacity | $800\,\frac{J}{kg \times K} - 1000\,\frac{J}{kg \times K}$ |
| Ti | Initial temperature | 150 °C – 200 °C |
| α | Biot coefficient | 0.5 – 0.7 |
| β | Thermal expansion coefficient | $10^{-6}\,\frac{1}{K} - 10^{-5}\,\frac{1}{K}$ |
| Tj | Injection temperature | 40 °C – 70 °C |

Figure 4(a1 & a2) displays the temperature distribution in the reservoir after one and 14 years respectively. The temperature drop near the injection well along the fracture network is clearly visible after one year from Figure 4(a1). More pathways for heat transport contribute to avoiding the thermal short-circuiting and delaying thermal breakthrough. It is observed that the pressurized zone around the injection well propagates more uniformly, rather than directionally, toward the production well. This result reveals that no preferential flow channels are formed between the injection well and the production well (Figure 4(a1 & a2)). The hot fluid occupying the pore space is gradually replaced by the injected cold fluid. With an increasing amount of the cold fluid in the reservoir, specific heat capacity of water decreases as shown in Figure 4(b1 & b2). The drop in specific heat capacity of water results in a reduced temperature at the production well. Figure 4(c1 & c2) shows that the increase in fluid viscosity assists in a stronger flow resistance within the cold zone.

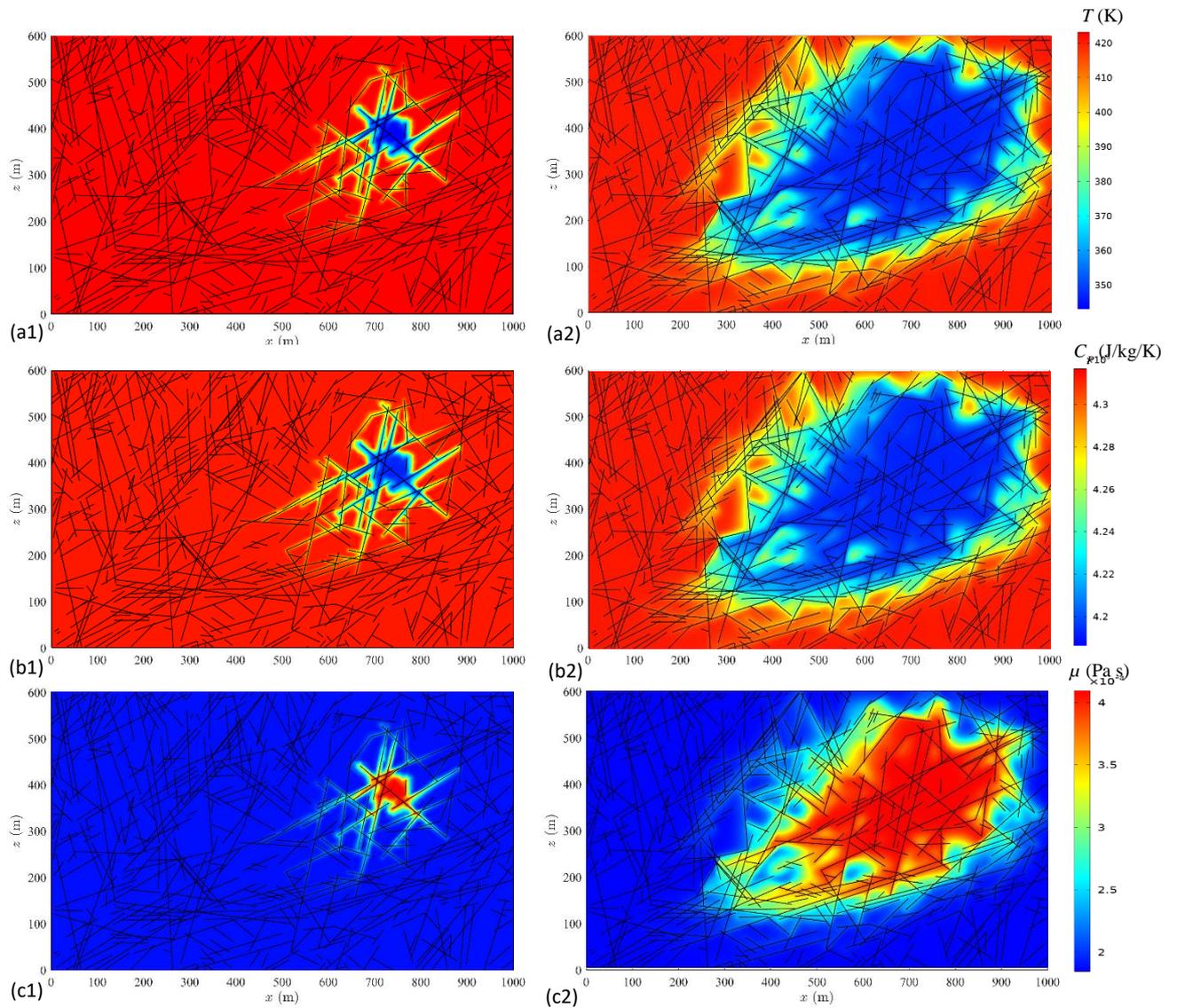

*Figure 4: (a) Fractured reservoir temperature, (b) specific heat capacity, and (c) dynamic fluid viscosity at (1) one and (2) 14 years. Here, the thermal breakthrough time at the production well is after 14 years.*

Thermoelasticity and poroelasticity are fully coupled in this study. Figure 5 shows the poroelastic stress distribution and Figure 6 displays the thermoelastic stress distribution in a fractured reservoir system. From Figure 5, it is clearly evident that a pressurized zone in the vicinity of the injection well develops as the fluid builds up in the wellbore. This is because within the early stage, the low permeable reservoir is insufficient to accommodate a continuous injection of fluid. The poroelastic stress is marginal when compared to thermoelastic stress as demonstrated in Figure 6 due to a large variation in temperature across the time. In other words, the influence of the poroelastic effect on the effective stress is limited, while thermoelasticity can affect the overall reservoir with the circulation of the cold fluid. In addition, the injection pressure gradually increases with time. An increasingly large pressurized zone moves towards the production well (see Figure 6(a2 & b2)). In Figure 6(a1 & b1) compressive stress due to the effect of strain compatibility (Koh et al. 2011) is visible for the region where the rock temperature drop is small.

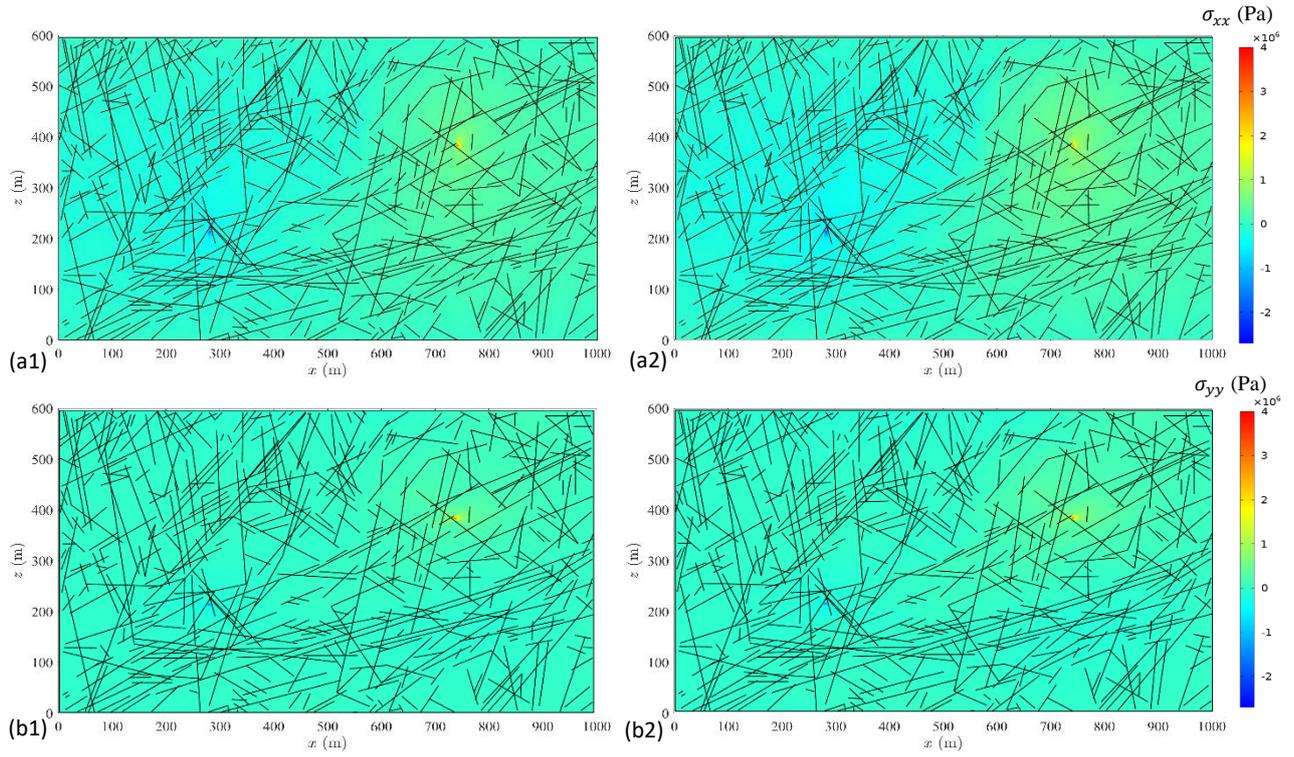

*Figure 5: Poroelastic stress distribution of (a) x-component and (b) y-component at time (1) one and (2) 14 years.*

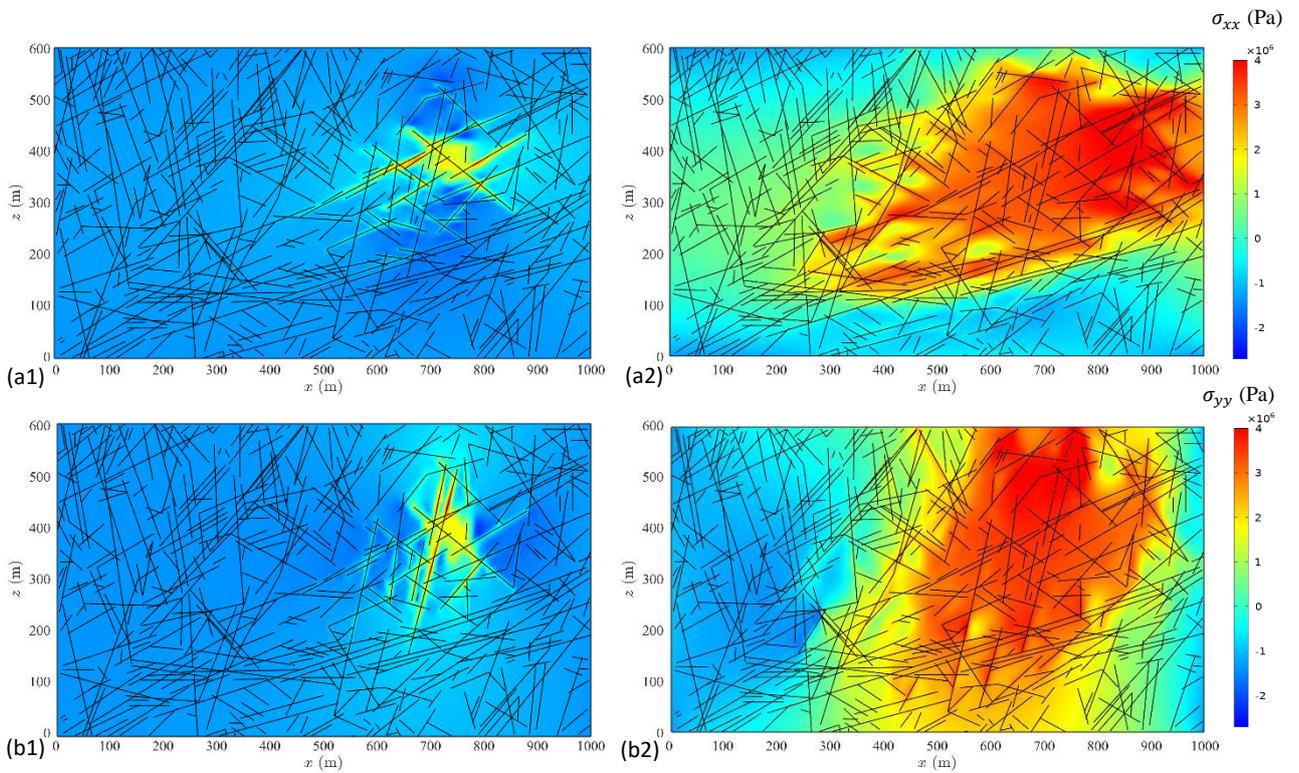

*Figure 6: Thermoelastic stress distribution of (a) x-component and (b) y-component at time (1) one and (2) 14 years.*

### 3.3. Plackett- Burman methodology

The design of experiments (DOE) methodology helps us to gather the maximum amount of information with the lowest number of simulations (Decaestecker et al. 2004, Mahmoodpour and Rostami 2017). Plackett-Burman methodology helps to identify the most important factors with as few simulations running as possible (Plackett and Burman 1946). Therefore, we can decide about important variables that require detailed measurements and negligible variables that we can use for correlations or some logical estimations. Therefore, we can save time and costs through eliminating unnecessary measurements. However, several factors are varied in each case, but each factor is examined independently. Plackett-Burman enables us to isolate the main effects of the factors from the interaction effects (Nijhuis et al. 1999; Yannis 2001). Results of the Plackett-Burman design are investigated based on the Hadamard matrices with a first degree polynomial equation as below:

$$R = c_0 + c_1 f_1 + \cdots + c_n f_n \qquad [15]$$

where R, c, and f show the estimated response, coefficient of factor, and factor respectively.

We needed at least 48 simulations to investigate the sensitivity analysis for 22 parameters as listed in Table 4. The details of the parameters with their respective symbols are listed in Table 5 and their corresponding minimum and maximum values are presented in Table 3. In order to determine which of the factors significantly affected the dependent variable of interest, an analysis of the variance (ANOVA) is performed. ANOVA presents the sum of squares (SS) which is used to estimate the factor main effects, the F-ratios (F) as the ratio of the respective mean-square effect (MS), and the mean-square error. The 'P'-probability values indicate the significant factors affecting the response (Vanaja and Shobha Rani 2007).

The Pareto chart shows the effects of each parameter in Figure 7 & 8. The dashed line divides the factors into two important and unimportant factors, based on the statistical analysis with some level of uncertainty (α). Here, $\alpha = 0.05$, which means that we are confident about the importance of the factors with 95% probability. It should be noted that this graph shows the importance of each parameter on the interested response. For the direction of the impact (positive or negative), the related coefficient value for each factor is shown by the positive or negative directions. Residual values corresponding to Figures 7 & 8, which show the difference between the estimated values and simulation results, are depicted in Figures 9 & 10. Most of the data fall onto a straight line which demonstrate that residuals show a distribution similar to the normal distribution (see Figures 11 & 12).

*Table 4: List of 48 simulated cases (shown in rows) for 22 input parameters (shown by columns) used for performing the sensitivity study. This table was developed by using the Plackett-Burman method. The minimum and maximum values for parameters listed in Table.*

| Case | E | ν | ρr | S1 | S2 | Pi | Pj | φr | kr | φf | f | Ap | σs | wr | λr | λf | Cr | Cf | Ti | α | β | Tj |
|---|---|---|---|---|---|---|---|---|---|---|---|---|---|---|---|---|---|---|---|---|---|---|
| 1 | 1 | -1 | -1 | -1 | 1 | 1 | -1 | 1 | 1 | -1 | -1 | 1 | -1 | -1 | 1 | 1 | 1 | -1 | 1 | -1 | 1 | -1 |
| 2 | 1 | -1 | -1 | -1 | -1 | 1 | -1 | -1 | -1 | -1 | 1 | 1 | -1 | 1 | -1 | 1 | -1 | -1 | -1 | 1 | 1 | -1 |
| 3 | -1 | -1 | 1 | -1 | -1 | 1 | 1 | 1 | -1 | 1 | -1 | 1 | -1 | -1 | 1 | 1 | 1 | -1 | 1 | 1 | 1 |
| 4 | -1 | -1 | -1 | -1 | 1 | -1 | -1 | -1 | -1 | 1 | 1 | -1 | 1 | -1 | 1 | -1 | -1 | -1 | 1 | 1 | -1 | 1 |
| 5 | -1 | -1 | 1 | 1 | 1 | -1 | 1 | -1 | 1 | -1 | -1 | 1 | 1 | 1 | 1 | -1 | 1 | 1 | 1 | 1 | 1 | -1 |
| 6 | 1 | 1 | -1 | -1 | 1 | -1 | -1 | 1 | 1 | 1 | -1 | 1 | -1 | 1 | -1 | -1 | 1 | 1 | 1 | 1 | -1 | 1 |
| 7 | -1 | -1 | -1 | -1 | 1 | 1 | -1 | 1 | -1 | 1 | -1 | -1 | -1 | 1 | 1 | -1 | 1 | 1 | -1 | -1 | 1 | -1 |
| 8 | 1 | 1 | 1 | 1 | -1 | 1 | 1 | 1 | 1 | 1 | -1 | -1 | -1 | 1 | -1 | -1 | -1 | -1 | 1 | 1 | -1 |
| 9 | -1 | 1 | -1 | -1 | 1 | 1 | 1 | -1 | 1 | -1 | 1 | -1 | -1 | 1 | 1 | 1 | 1 | -1 | 1 | 1 | 1 | 1 |
| 10 | -1 | 1 | 1 | -1 | 1 | -1 | 1 | -1 | -1 | -1 | 1 | 1 | -1 | 1 | 1 | -1 | -1 | 1 | -1 | -1 | 1 | 1 |
| 11 | -1 | 1 | 1 | 1 | -1 | 1 | -1 | 1 | -1 | -1 | 1 | 1 | 1 | -1 | 1 | 1 | 1 | 1 | 1 | -1 | -1 |
| 12 | 1 | 1 | -1 | -1 | -1 | -1 | 1 | -1 | -1 | -1 | -1 | 1 | 1 | -1 | 1 | -1 | 1 | -1 | -1 | -1 | 1 | 1 |

| | | | | | | | | | | | | | | | | | | | | |
|---|---|---|---|---|---|---|---|---|---|---|---|---|---|---|---|---|---|---|---|---|
| 13 | 1 | 1 | -1 | 1 | -1 | 1 | -1 | -1 | 1 | 1 | 1 | 1 | -1 | 1 | 1 | 1 | 1 | 1 | -1 | -1 | -1 | -1 |
| 14 | 1 | 1 | 1 | 1 | 1 | -1 | -1 | -1 | -1 | 1 | -1 | -1 | -1 | -1 | 1 | 1 | -1 | 1 | -1 | 1 | -1 | -1 |
| 15 | 1 | -1 | -1 | 1 | 1 | 1 | 1 | -1 | 1 | 1 | 1 | 1 | 1 | -1 | -1 | -1 | -1 | 1 | -1 | -1 | -1 | -1 |
| 16 | -1 | 1 | -1 | 1 | -1 | -1 | -1 | 1 | 1 | -1 | 1 | 1 | -1 | -1 | 1 | -1 | -1 | 1 | 1 | 1 | -1 | 1 |
| 17 | 1 | 1 | -1 | 1 | 1 | 1 | -1 | -1 | 1 | -1 | -1 | 1 | 1 | 1 | -1 | 1 | -1 | 1 | -1 | -1 | 1 | 1 | 1 |
| 18 | 1 | -1 | 1 | -1 | 1 | -1 | -1 | -1 | 1 | 1 | -1 | 1 | 1 | -1 | -1 | 1 | -1 | -1 | 1 | 1 | 1 | -1 |
| 19 | -1 | -1 | 1 | 1 | -1 | 1 | -1 | 1 | -1 | -1 | 1 | 1 | 1 | -1 | 1 | 1 | -1 | -1 | 1 | -1 | -1 | 1 |
| 20 | -1 | 1 | 1 | -1 | -1 | 1 | -1 | -1 | 1 | 1 | 1 | -1 | 1 | -1 | 1 | -1 | -1 | 1 | 1 | 1 | 1 | -1 |
| 21 | 1 | -1 | 1 | -1 | -1 | -1 | 1 | 1 | -1 | 1 | 1 | -1 | -1 | 1 | -1 | -1 | 1 | 1 | 1 | -1 | 1 | -1 |
| 22 | 1 | -1 | -1 | 1 | 1 | 1 | -1 | 1 | -1 | 1 | -1 | -1 | 1 | 1 | 1 | -1 | 1 | 1 | 1 | 1 | 1 | 1 |
| 23 | -1 | 1 | 1 | 1 | 1 | -1 | 1 | 1 | 1 | 1 | 1 | -1 | -1 | -1 | -1 | 1 | -1 | -1 | -1 | -1 | 1 | 1 |
| 24 | 1 | 1 | 1 | -1 | 1 | -1 | 1 | -1 | -1 | 1 | 1 | 1 | -1 | 1 | 1 | 1 | 1 | 1 | -1 | -1 | -1 |
| 25 | 1 | 1 | 1 | -1 | -1 | -1 | -1 | 1 | -1 | -1 | -1 | -1 | 1 | 1 | -1 | 1 | -1 | 1 | -1 | -1 | -1 | 1 |
| 26 | 1 | -1 | 1 | 1 | 1 | 1 | 1 | -1 | -1 | -1 | -1 | 1 | -1 | -1 | -1 | -1 | 1 | 1 | -1 | 1 | -1 | 1 |
| 27 | 1 | -1 | 1 | 1 | -1 | -1 | 1 | -1 | -1 | 1 | 1 | 1 | -1 | 1 | -1 | 1 | -1 | -1 | 1 | 1 | 1 | 1 |
| 28 | -1 | -1 | 1 | 1 | 1 | 1 | -1 | 1 | 1 | 1 | 1 | 1 | -1 | -1 | -1 | -1 | 1 | -1 | -1 | -1 | -1 | 1 |
| 29 | -1 | -1 | -1 | -1 | -1 | -1 | -1 | -1 | -1 | -1 | -1 | -1 | -1 | -1 | -1 | -1 | -1 | -1 | -1 | -1 | -1 | -1 |
| 30 | 1 | 1 | -1 | 1 | 1 | 1 | 1 | 1 | -1 | -1 | -1 | -1 | 1 | -1 | -1 | -1 | -1 | 1 | 1 | -1 | 1 | -1 |
| 31 | -1 | -1 | -1 | 1 | 1 | -1 | 1 | -1 | 1 | -1 | -1 | -1 | 1 | 1 | -1 | 1 | 1 | -1 | -1 | 1 | -1 | -1 |
| 32 | -1 | -1 | 1 | 1 | -1 | 1 | 1 | -1 | -1 | 1 | -1 | -1 | 1 | 1 | 1 | -1 | 1 | -1 | 1 | -1 | -1 | 1 |
| 33 | -1 | -1 | -1 | 1 | -1 | -1 | -1 | -1 | 1 | 1 | -1 | 1 | -1 | 1 | -1 | -1 | -1 | 1 | 1 | -1 | 1 | 1 |
| 34 | -1 | 1 | 1 | 1 | 1 | 1 | -1 | -1 | -1 | -1 | 1 | -1 | -1 | -1 | -1 | 1 | 1 | -1 | 1 | -1 | 1 | -1 |
| 35 | 1 | 1 | 1 | 1 | -1 | -1 | -1 | -1 | 1 | -1 | -1 | -1 | -1 | 1 | 1 | -1 | 1 | -1 | 1 | -1 | -1 | -1 |
| 36 | 1 | -1 | 1 | -1 | 1 | -1 | -1 | 1 | 1 | 1 | 1 | -1 | 1 | 1 | 1 | 1 | 1 | -1 | -1 | -1 | -1 | 1 |
| 37 | 1 | -1 | 1 | -1 | -1 | 1 | 1 | 1 | 1 | -1 | 1 | 1 | 1 | 1 | -1 | -1 | -1 | -1 | 1 | -1 | 1 | -1 | -1 |
| 38 | -1 | -1 | -1 | 1 | 1 | -1 | 1 | 1 | -1 | -1 | 1 | -1 | -1 | 1 | 1 | -1 | 1 | -1 | 1 | -1 | -1 |
| 39 | -1 | 1 | 1 | -1 | 1 | 1 | -1 | -1 | 1 | -1 | -1 | 1 | 1 | 1 | -1 | 1 | -1 | 1 | -1 | -1 | 1 | 1 |
| 40 | -1 | -1 | 1 | -1 | -1 | -1 | -1 | 1 | 1 | -1 | 1 | -1 | 1 | -1 | -1 | 1 | 1 | -1 | 1 | 1 | -1 |
| 41 | -1 | 1 | -1 | 1 | -1 | -1 | 1 | 1 | 1 | 1 | -1 | 1 | 1 | 1 | 1 | -1 | -1 | -1 | -1 | 1 | -1 |
| 42 | 1 | 1 | 1 | -1 | 1 | 1 | 1 | 1 | 1 | -1 | -1 | -1 | -1 | 1 | -1 | -1 | -1 | -1 | 1 | 1 | -1 | 1 |
| 43 | -1 | 1 | -1 | -1 | -1 | -1 | 1 | 1 | -1 | 1 | -1 | 1 | -1 | -1 | -1 | 1 | 1 | -1 | 1 | 1 | -1 | -1 |
| 44 | 1 | -1 | -1 | -1 | -1 | 1 | 1 | -1 | 1 | -1 | 1 | -1 | -1 | -1 | 1 | 1 | -1 | 1 | 1 | -1 | -1 | 1 |
| 45 | -1 | 1 | -1 | -1 | -1 | 1 | 1 | -1 | 1 | 1 | -1 | -1 | 1 | -1 | -1 | 1 | 1 | 1 | -1 | 1 | -1 | 1 |
| 46 | 1 | 1 | -1 | 1 | -1 | 1 | -1 | -1 | -1 | 1 | 1 | -1 | 1 | 1 | -1 | -1 | 1 | -1 | -1 | 1 | 1 | 1 |
| 47 | -1 | 1 | -1 | -1 | 1 | 1 | 1 | 1 | -1 | 1 | 1 | 1 | 1 | -1 | -1 | -1 | -1 | 1 | -1 | -1 | -1 |
| 48 | 1 | -1 | -1 | 1 | -1 | -1 | 1 | 1 | 1 | -1 | 1 | -1 | 1 | -1 | -1 | 1 | 1 | 1 | 1 | -1 | 1 | 1 |

Table 5: Weighted coefficients and P-values for each parameters used in the T10 case.

| Parameter | Term | Symbol | Coefficient | P-Value |
|---|---|---|---|---|
| | | | 3824 | 0.000 |
| Young's modulus | E | A | 17 | 0.932 |
| Poisson's ratio | PR | B | 213 | 0.292 |
| Rock density | RD | C | 76 | 0.703 |
| Horizontal stress | S1 | D | 213 | 0.291 |
| Vertical stress | S2 | E | 85 | 0.670 |

| Initial pressure | IP | F | 794 | 0.000 |
|---|---|---|---|---|
| Injection pressure | InjP | G | -933 | 0.000 |
| Rock porosity | RP | H | 69 | 0.729 |
| Rock permeability | Rper | J | -1183 | 0.000 |
| Fracture porosity | FP | K | 111 | 0.578 |
| Fracture roughness | FR | L | 572 | 0.008 |
| Fracture aperture | FA | M | -1739 | 0.000 |
| Closure stress | ES | N | -201 | 0.319 |
| Wellbore radius | WR | O | 82 | 0.682 |
| Rock thermal conductivity | RT | P | -186 | 0.355 |
| Fracture thermal conductivity | FT | Q | -158 | 0.431 |
| Rock specific heat capacity | RSH | R | 549 | 0.010 |
| Fracture specific heat capacity | FSH | S | 45 | 0.822 |
| Initial temperature | IT | T | -220 | 0.276 |
| Biot coefficient | BC | U | -77 | 0.698 |
| Thermal expansion coefficient | TE | V | 1060 | 0.000 |
| Injection temperature | InjT | W | -661 | 0.003 |

### 3.4. Sensitivity for thermal breakthrough, mass flow rate, and total energy recovery

We performed the sensitivity analysis using the Plackett-Burman methodology in two stages. First, we developed the sensitivity analysis for the thermal breakthrough time at the production well for a temperature drop of 10 ℃, 20 ℃, and 30 ℃. Based on these results, we further identified the sensitive parameters for mass flow rate from the production well and the overall energy recovery for 10 ℃, 20 ℃, and 30 ℃ temperature drop at the production well. In our next study, we have investigated the sensitivity study for the thermal breakthrough at the production well for a 10% and 20% temperature drop. Similarly, we identified the critical parameters for the mass flow rate through the production well and the overall energy recovery for a 10% and 20% temperature at the production well.

Figure 7(a1) shows the Pareto chart results for the sensitivity in a thermal breakthrough time at the production well for a temperature drop by 10 ℃. The corresponding residual and histogram plots are demonstrated in Figures 9(a1) and Figure 11(a1). Based on our Plackett-Burman methodology, we found that the most important parameter controlling the thermal breakthrough time is the fracture aperture, followed by rock matrix permeability (see Figure 7(a1, b1 & c1) ). However, both of these parameters impact negatively on the breakthrough time. From the sensitivity study of mass flow rate, we found that rock permeability and wellbore radius are the most important parameters and impact positively (see Figure 7(a2, b2 & c2)). Therefore, higher mass flux causes faster thermal depletion, thus inverse behavior of rock permeability on temperature breakthrough and mass flux is obvious. For the M30 case, only the rock matrix permeability is the important parameter, due to the fact that a major pathway develops through the fractures in the early time for all cases and the thermal effects govern the reservoir mechanical properties. However, in later time, the thermal impact becomes recessive and the overall mass flux reaches a quasi-steady-state. Figure 7(a3, b3 & c3) shows the sensitivity for energy recovery from the production well. Here, the most important parameter is the wellbore radius and fracture aperture, where the former parameter enhances the energy recovery and the latter parameter reduces it. From Figures 7(c1) and 7(c3), it is clearly evident that the weightage of wellbore radius and fracture aperture is approximately the same as for the 10 ℃ case (E10), whereas wellbore radius becomes almost twice as important than the fracture aperture for the 30 ℃ case (E30). This is due to the fact that delayed thermal breakthrough time reduces mass flux through the fractures caused by increased viscosity of water.

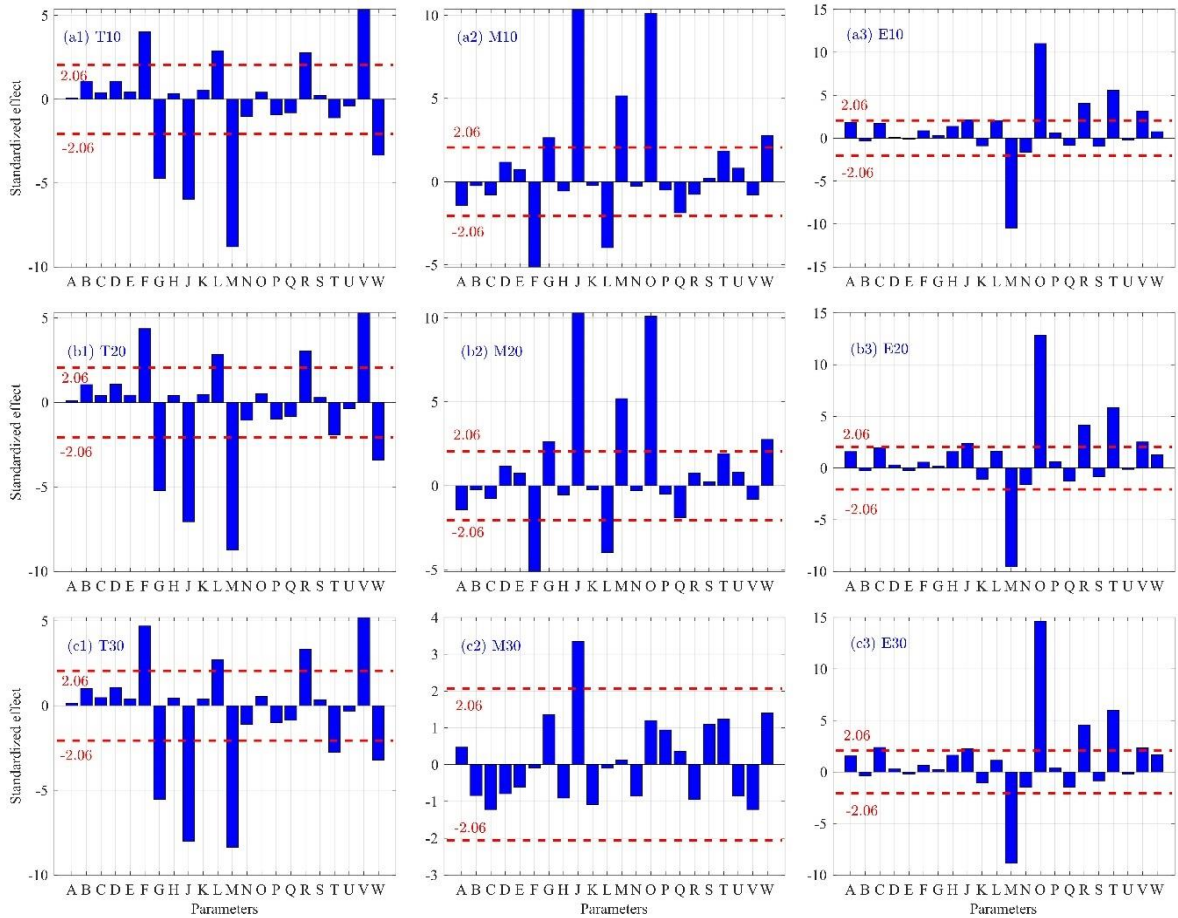

*Figure 7: Pareto charts of 22 parameters to investigate the comparative impact on THM model for the production well breakthrough temperature. The list of 22 parameters is given in Table 5 with their corresponding symbols. The left column represent the temperature breakthrough sensitivity for a (a1) 10 ℃, (b1) 20 ℃, and (c1) 30 ℃ temperature drop, the middle column shows the sensitivity in the mass flow rate at the production well with a (a2) 10 ℃, (b2) 20 ℃, and (c2) 30 ℃ drop in the well temperature, and the right column indicates the energy sensitivity for corresponding the temperature drop by (a3) 10 ℃, (b3) 20 ℃, and (c3) 30 ℃. Here, T10 means the temperature sensitivity study for a 10 ℃ temperature drop. Similarly, M10 and E10 indicates a sensitivity for a temperature drop at the production well for 10 ℃, respectively.*

Numerical simulation results will be the same for a 10% & 20% drop in temperature at the production well for 150 ℃ and a 10% drop for 200 ℃ in initial reservoir temperature. However, the Pareto chart with over 48 simulations gives different sensitivity results, e.g., as shown in Figures 7(b1-b3) and 8(a1-a3), due to a wide sample space. For cases with a 10% and 20% temperature drop, the thermal breakthrough time (T10% case, see Figure 8(a1)) and energy recovery sensitivity (E10% case, see Figure 8(a3)) results show similar outcomes when compared to the T20 & E20 cases. The disparity in the mass flow sensitivity result (M10% case, see Figure 8(a2)) is prominent when compared with M20 (see Figure 7(b2)) cases. The sensitivity trend and the crucial parameters are respectively same for all temperature drops (10℃ - 30 ℃ & 10% - 20%). However, mass flow rate does not follow this trend. The most important parameters for the M10% case are fracture aperture and initial pressure, whereas for the M20% case, it is fracture aperture, initial pressure, fracture roughness, rock permeability, and wellbore radius. The higher sensitivity of the M20% case is due to a slightly increased relative importance of the parameter from the M10% case.

The absolute value of residual between numerical simulation and the regression analysis for all 48 cases are shown in Figures 9 and 10. It is clearly visible that for most of the cases, the residual value falls near the zero line indicating a normal distribution also demonstrated by Figures 11 and 12. Case number 34 from Table 4 in T10, T20 & T30, M10, M20 & M30, and case number 21 in E10, E20 & E30 have higher residual values due to a extreme range of input parameters listed in Tables 4 and 5.

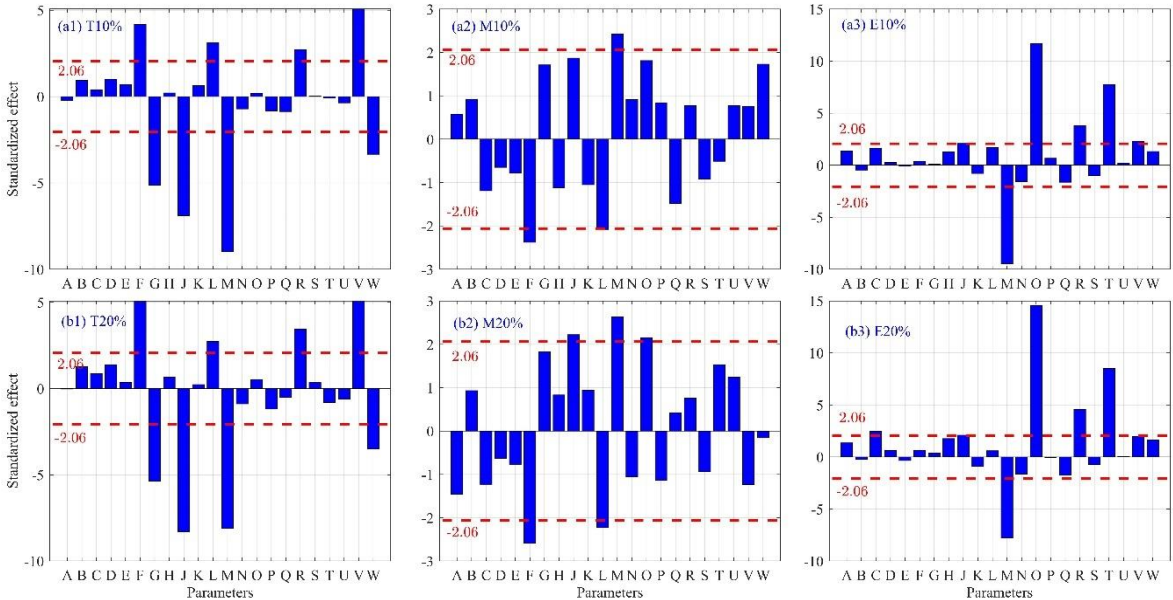

*Figure 8: Pareto charts of 22 parameters to investigate the comparative impact on THM model for the production well breakthrough temperature. The list of 22 parameters is given in Table 5. The left column represents the temperature breakthrough sensitivity for a (a1) 10% and (b1) 20% temperature drop, the middle column shows the sensitivity in the mass flow rate at the production well for a (a2) 10% and (b2) 20% drop in the well temperature, and the right column indicates the energy sensitivity for the corresponding temperature drop by (a3) 10% and (b3) 20%.*

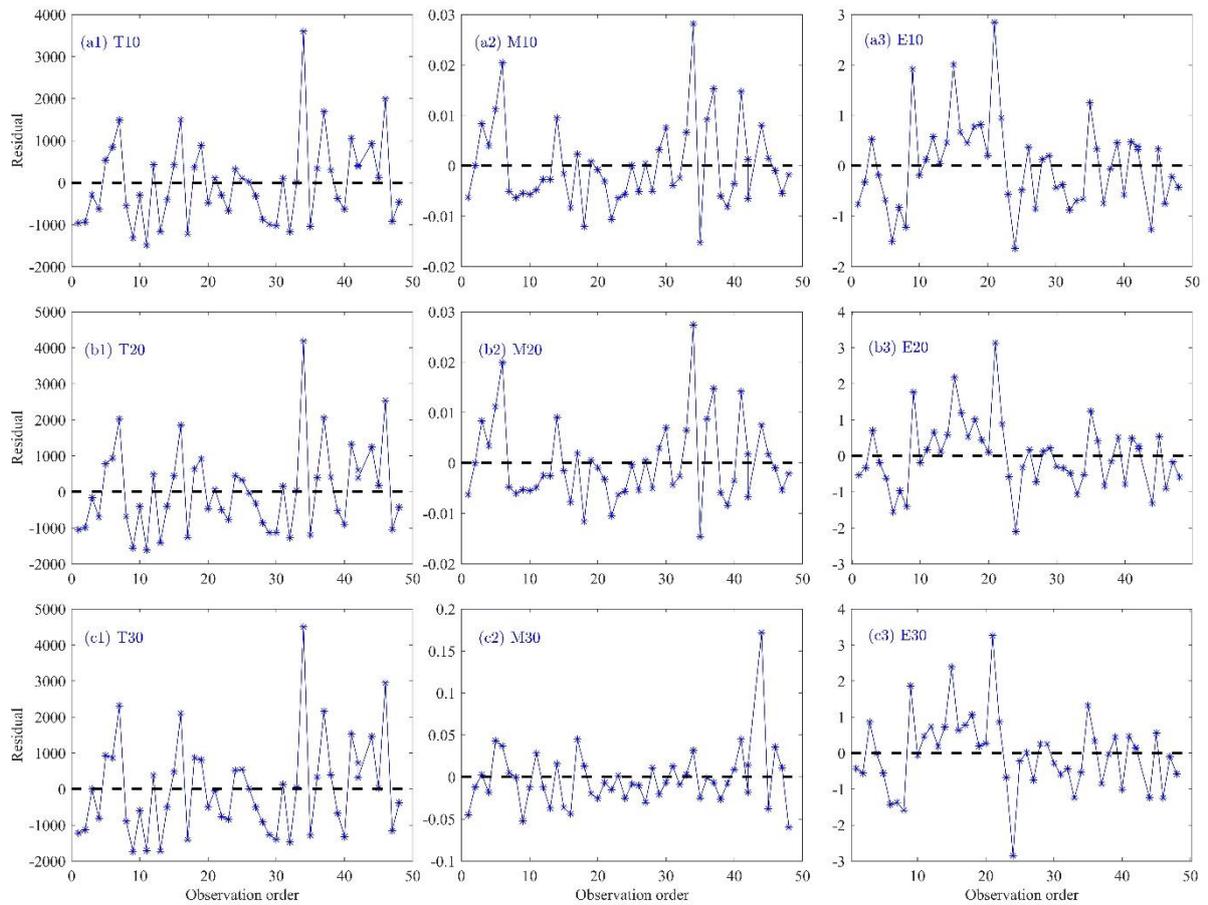

*Figure 9: Residual between simulated and regression results for 48 cases. The left column represents the temperature breakthrough sensitivity for a (a1) 10 ℃, (b1) 20 ℃, and (c1) 30 ℃ temperature drop, the middle column shows the sensitivity in the mass flow rate at the production well for a (a2) 10 ℃, (b2) 20 ℃, and (c2) 30 ℃ drop in the well temperature, and the right column indicates the energy sensitivity for the corresponding temperature drop by (a3) 10 ℃, (b3) 20 ℃, and (c3) 30 ℃.*

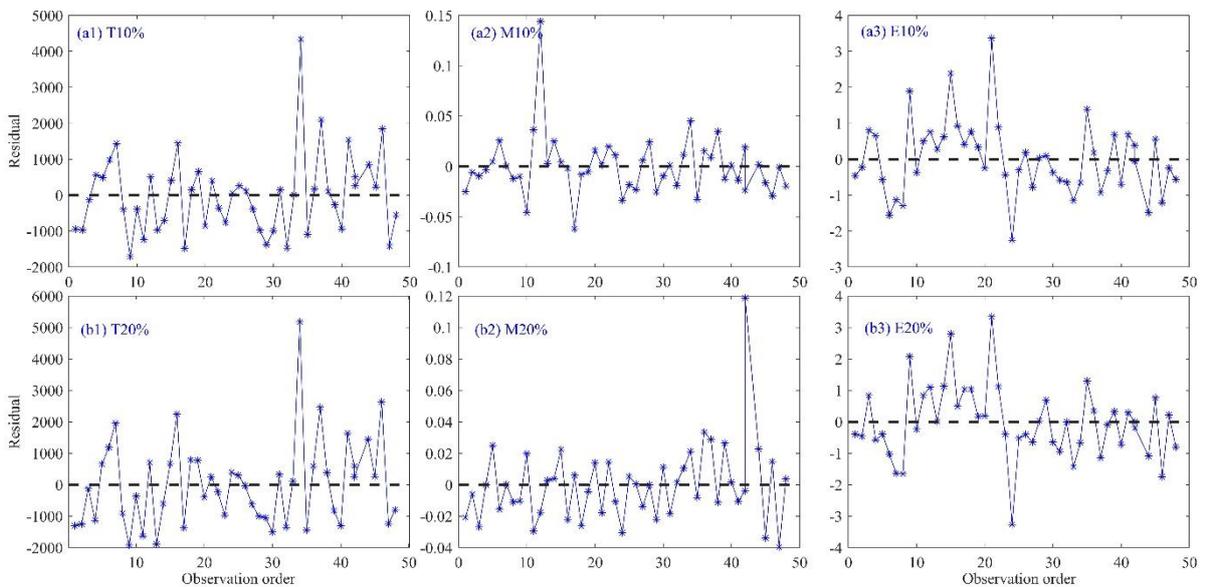

*Figure 10: Residual between simulated and regression results for 48 cases. The left column represents the temperature breakthrough sensitivity for a (a1) 10% and (b1) 20% temperature drop, the middle column shows the sensitivity in the mass flow rate at the production well for a (a2) 10% and (b2) 20% drop in the well temperature, and the right column indicates the energy sensitivity for the corresponding temperature drop by (a3) 10% and (b3) 20%.*

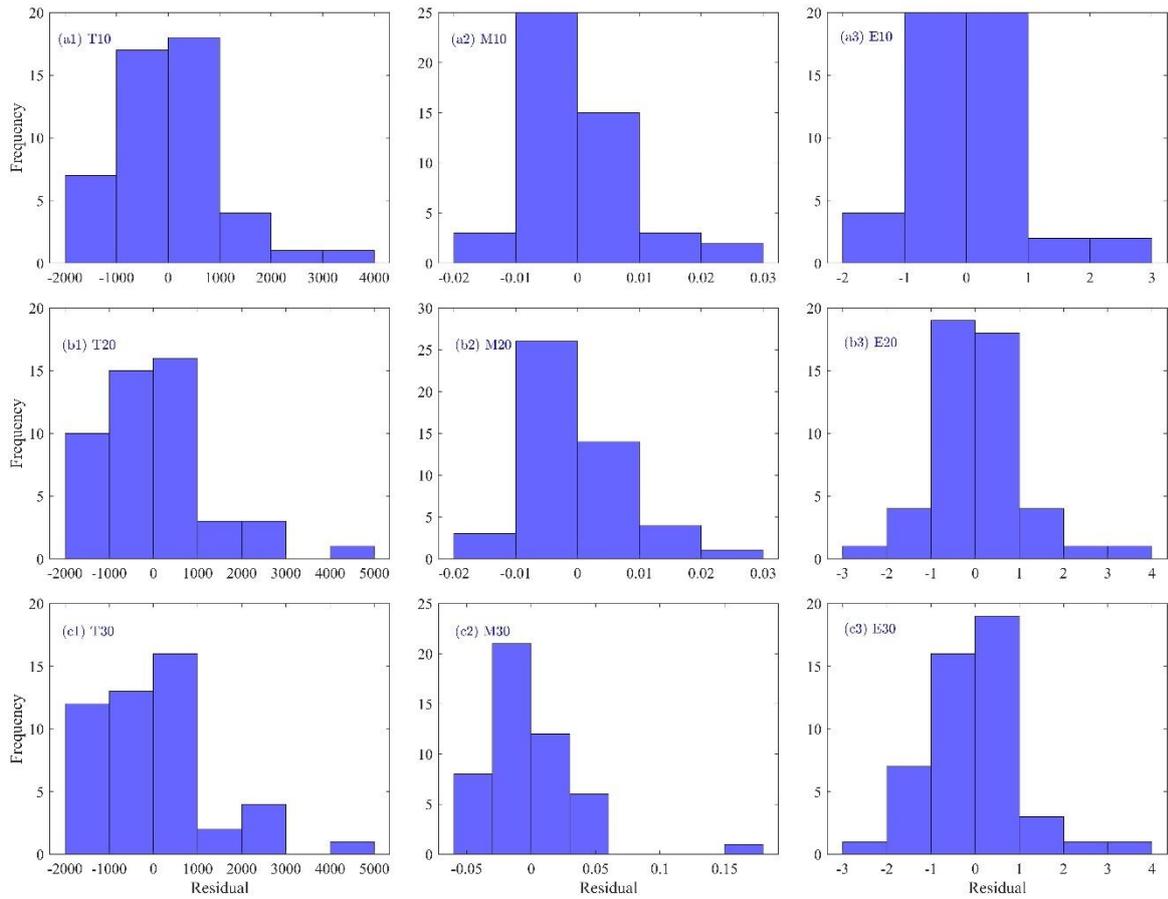

*Figure 11: Histogram of residuals between simulated and regression results. The left column represents the temperature breakthrough sensitivity for a (a1) 10 ℃, (b1) 20 ℃, and (c1) 30 ℃ temperature drop, the middle column shows the sensitivity in the mass flow rate at the production well for a (a2) 10 ℃, (b2) 20 ℃, and (c2) 30 ℃ drop in the well temperature, and the right column indicates the energy sensitivity for the corresponding temperature drop by (a3) 10 ℃, (b3) 20 ℃, and (c3) 30 ℃.*

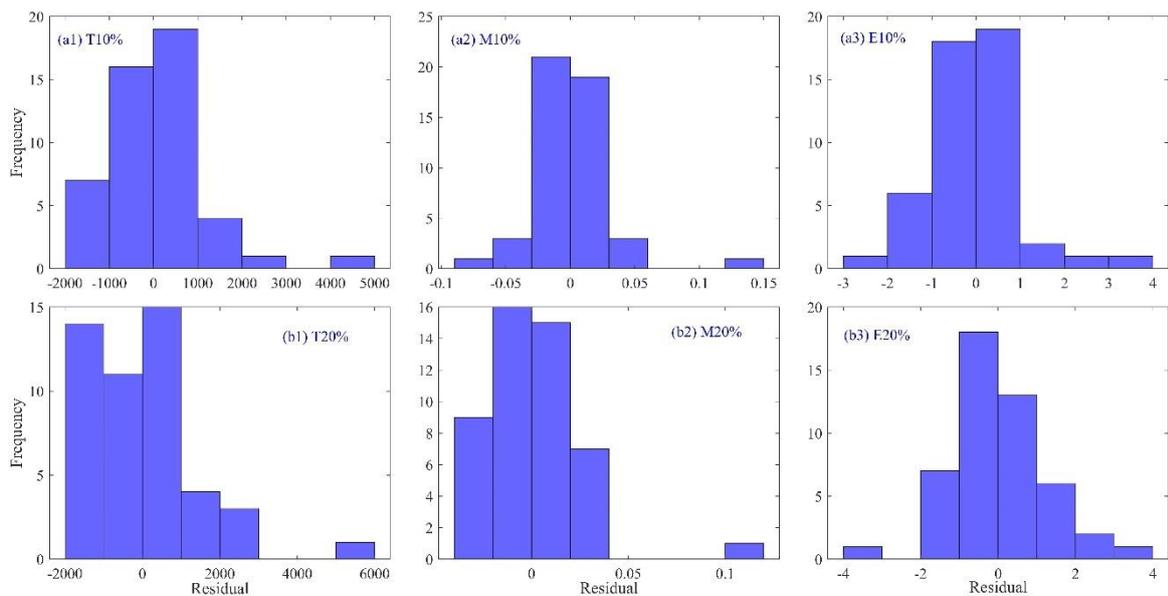

*Figure 12: Histogram of residuals between simulated and regression results for 48 cases. The left column represents the temperature breakthrough sensitivity for a (a1) 10% and (b1) 20% temperature drop, the middle column shows the sensitivity in the mass flow rate at the production well for a (a2) 10% and (b2) 20% drop in the well temperature, and right column indicates the energy sensitivity for the corresponding temperature drop by (a3) 10% and (b3) 20%.*

# 4. Conclusions

We have successfully demonstrated the THM mechanisms involved in EGS operation in a two-dimensional discretely fractured reservoir. We have presented a fully coupled poroelastic and thermoelastic approach to investigate stress development during EGS operation. Further, using the DOE method or the Plackett-Burman method, we have performed sensitivity analyses to identify the most critical parameters influencing the THM process. Within the framework of MEET H2020 EU project, we have presented a detailed and comprehensive sensitivity analysis. Based on our Plackett-Burman method, we identified 22 parameters involved in THM mechanism and performed 48 fully coupled compositional numerical simulations. We fixed three major outcomes to investigate the sensitivity study: the thermal breakthrough time, the mass flow rate, and the total energy extraction. We observed that the residual between the numerical and regression analysis for all three outcomes, at five different temperature reductions (10 ℃, 20 ℃ & 30 ℃, and 10% & 20%), are floating around zero line and follow a normal distribution. Therefore, we can conclude that our existing sensitivity analysis comprehensively estimated the relative importance of individual parameters. Based on our sensitivity study, we found that fracture aperture, rock matrix permeability, and wellbore radius are the three critical parameters controlling the overall THM process.

**Acknowledgement**

The work is conducted as a part of the MEET project that has received funding from the European Union's Horizon 2020 research and innovation programme under grant agreement No 792037. Authors have received support from the Group of Geothermal Science and Technology, Institute of Applied Geosciences, Technische Universität Darmstadt.